\documentclass[prl,twocolumn,amsmath,amssymb]{revtex4}

\usepackage{color}

\newcommand{\beq}{\begin{equation}}
\newcommand{\eeq}{\end{equation}}

\newcommand{\asin}{\text{asin}\,}

\begin{document}

\title{Bounding the set of quantum correlations}

\author{Miguel Navascu\'{e}s}
%\email{Miguel.Navascues@icfo.es}
\author{Stefano Pironio}
%\email{Stefano.Pironio@icfo.es}
\author{Antonio Ac\'{\i}n}
%\email{Antonio.Acin@icfo.es}
\affiliation{ICFO-Institut de Ciencies Fotoniques, Mediterranean Technology Park, 08860 Castelldefels (Barcelona), Spain\\
E-mail: \emph{\texttt{Miguel.Navascues@icfo.es}, \texttt{Stefano.Pironio@icfo.es}, \texttt{Antonio.Acin@icfo.es}}}

\date{July 18, 2006}

\begin{abstract}
We introduce a hierarchy of conditions necessarily satisfied by
any distribution $P_{\alpha\beta}$ representing the probabilities
for two separate observers to obtain outcomes $\alpha$ and $\beta$
when making local measurements on a shared quantum state. Each
condition in this hierarchy is formulated as a semidefinite
program. Among other applications, our approach can be used to obtain upper-bounds on the
quantum violation of an arbitrary Bell inequality. It yields, for
instance, tight bounds for the violations of the Collins~{et al.}
inequalities.
\end{abstract}

\maketitle

The correlations between two separated physical systems can be
characterized by the joint probabilities $P_{\alpha\beta}$ that an
observer who performs a measurement $X$ on the first system gets
an outcome $\alpha\in X$ and that an observer making a measurement
$Y$ on the second system gets an outcome $\beta\in Y$. If the
observed system is in an entangled state, these joint
probabilities may violate a Bell inequality, implying that quantum
theory is not, in Bell's terminology, locally causal \cite{bell}.
Although quantum correlations are not constrained by Bell's
locality principle, they are not arbitrary since a general joint
distribution $P_{\alpha\beta}$ cannot always be viewed as
originating from measurements performed on a shared quantum system
\cite{kt}.

In this paper, we investigate the restrictions on bipartite
correlations imposed by the quantum formalism. The question that
we seek to answer is the following: Given an arbitrary
distribution $P_{\alpha\beta}$, do there exist a quantum state
$\rho$ on a joint Hilbert space
$\mathcal{H}_A\otimes\mathcal{H}_B$ and local measurement
operators $E_{\alpha}=\tilde{E}_{\alpha}\otimes{I}$ and
$E_{\beta}=I\otimes \tilde{E}_{\beta}$, such that $
P_{\alpha\beta}=\text{tr}\left(E_{\alpha}E_{\beta}\,\rho\right) $?

From a fundamental point of view, one motivation for studying this problem is simply to understand which correlations can arise between two systems within our current description of nature. Another is to develop tools to detect the possible non-quantumness of some set of empirically observed correlations. Answering the above question is also of practical interest for various applications in quantum science, for instance, for the design of nonlocality tests more resistant to experimental noise and detector inefficiencies. In general, characterizing the set of quantum correlations is essential to understand better the extent to which quantum mechanics is useful in information processing tasks such as communication complexity \cite{cleve} and key distribution \cite{barrett,acin2}. An usual problem in these contexts is to determine what is the maximal violation of a Bell inequality allowed by quantum mechanics.

Answering the above question is in general a difficult task; the simple strategy of searching over all quantum states $\rho$ and measurement operators $E_\mu$, which in principle can be of
arbitrary dimension, is clearly unfeasible. The first to tackle this problem  was Tsirelson \cite{tsir}, who derived the maximal quantum violation for the Clauser-Horne-Shimony-Holt (CHSH) inequality \cite{chsh}. Following his seminal work, many attempts have been made to understand better how the quantum formalism constrains the correlations between two parties \cite{bm}, but most of the obtained results only apply to a restricted set of situations or lack an efficient generalization. Among these, we mention the works of Landau \cite{landau} and Wehner \cite{wehner} who realized that the problem of deciding if a set of ``correlation functions'' obtained from measurements with binary outcomes admits a quantum representation can be cast as a semidefinite program (SDP). Their approach is interesting in that semidefinite programs are convex optimization problems for which powerful computational and theoretical methods have been developed \cite{sdp}.

We show here how to design criteria that distinguish correlations that can be reproduced through local measurements on a quantum state from those which cannot. Testing any one of these criteria amounts to solve a SDP. Contrary to previous constructions, our method can be applied to any distribution $P_{\alpha\beta}$ of joint probabilities, corresponding to a configuration with an arbitrary number of measurement choices and outcomes. Seen as a
whole, the tests that we introduce exhibit a hierarchical structure,
in the sense that they can be organized as a sequence of conditions, each
condition in the sequence being stronger than the preceding one.
As an illustration of the effectiveness of our approach, we present two applications of it. First, we derive a
non-linear inequality satisfied by quantum mechanics which
strengthens a previous inequality due to Tsirelson \cite{tsir2}, Landau
\cite{landau} and Masanes~\cite{masanes}. As a second application,
we give a tight bound for the violation of the Collins et al.
inequalities~\cite{cglmp}.

\paragraph{Preliminaries.}
Before entering in the details of our construction, let us first
give some definitions and state the assumptions made through
this paper. We assume that the two parties, Alice and Bob, choose
their measurements from a finite set of possibilities, and that
each measurement may yield one out of a finite set of outcomes.
Note that we think of outcomes corresponding to different
measurements as being labeled distinctly, so that each outcome
$\alpha$ of Alice (or $\beta$ of Bob) is unambiguously associated to
a unique measurement $X$ (or $Y$).

Refining the statement made earlier, we say that a distribution
$P_{\alpha\beta}$ admits a quantum representation if there exist a
joint quantum state $\rho$ on $\mathcal{H}_A\otimes\mathcal{H}_B$,
a set of \emph{projection} operators
$E_{\alpha}=\tilde{E}_{\alpha}\otimes{I}$ acting on Alice's system
and a set of \emph{projection} operators $E_{\beta}=I\otimes
\tilde{E}_{\beta}$ acting on Bob's system, such that
\begin{equation}\label{qprob}
P_{\alpha\beta}=\text{tr}\left(E_{\alpha}E_{\beta}\,\rho\right)\,.
\end{equation}
Projectors corresponding to outcomes belonging to the same
measurement $M$ should (\emph{i}) be orthogonal: $E_\mu E_\nu=0$
for $\mu,\nu\in M$, $\mu\neq \nu$, and (\emph{ii}) sum to the
identity: $\sum_{\mu\in M} E_\mu=I$. By definition, we also have
that (\emph{iii}) $E_\mu^2=E_\mu^\dag=E_\mu$ and that (\emph{iv})
the projectors on Alice and Bob's side commute with each other:
$[E_\alpha,E_\beta]=0$.

Note that the most general description of a quantum
measurement corresponds to a positive operator valued measure
(POVM) rather than a set of projection operators. But
since a POVM can be viewed as a projective measurement on a system
of larger dimension, and since we do not impose any constraints on
the dimension of the Hilbert space
$\mathcal{H}_A\otimes\mathcal{H}_B$, no generality is lost with
our definition.

\paragraph{Necessary conditions for quantum probabilities.}
We now introduce a family of conditions satisfied by any quantum
distribution $P_{\alpha\beta}$. We thus start by assuming that
there is a quantum state $\rho$ and a set $\{E_\mu\}$ of
projection operators satisfying Eq.~\eqref{qprob} and
properties (\emph{i})-(\emph{iv}), and seek new implications from
these assumptions.

By taking products of the operators $E_\mu$ and linear
combinations of such products, we can define new operators, for
instance $E_\alpha E_\beta E_{\alpha'}$ or $\sum_{\alpha}c_\alpha
E_\alpha$ (note that these new operators are not necessarily
projection operators anymore, nor even hermitian operators). Let
$\mathcal{S}=\{S_1,\ldots,S_n\}$ be a set of $n$ such operators.
Associate to the set $\mathcal{S}$ a $n\times n$ matrix $\Gamma$
through \beq\label{gamma} \Gamma_{ij}=\text{tr}\left(S_i^\dag
S_j\,\rho\right)\,. \eeq By construction, the matrix $\Gamma$ is
hermitian, it satisfies the linear identities
\begin{gather}
\label{lin1}\sum_{i,j} c_{ij} \Gamma_{ij}=0 \qquad \text{if } \sum_{i,j} c_{ij} S^\dag_iS_j=0\,,\hspace{1.9cm}
\displaybreak[0]
\\
\label{lin2}\begin{split}\sum_{i,j} c_{ij} \Gamma_{ij}=\sum_{\alpha,\beta}&d_{\alpha\beta}P_{\alpha\beta}\\ &\text{if } \sum_{i,j}c_{ij} S_i^\dag S_j=\sum_{\alpha,\beta} d_{\alpha\beta} E_\alpha E_\beta\,,\hspace{0.5cm}\end{split}
\end{gather}
and it is positive semidefinite, \beq\label{sdpcond} \Gamma\succeq
0\,. \eeq The linear constraints \eqref{lin1} directly follow from
the linearity of the trace in \eqref{gamma}. The important point
is that they partly reflect the properties \emph{(i)-(iv)}
satisfied by the operators $E_\mu$. For instance suppose that
$\mathcal{S}$ contains an operator $S_{j}=E_\mu$ and a subset of
operators $\{S_k\mid k\in\mathcal{K}\}=\{E_\mu E_\nu\mid \nu \in
M\}$ for some measurement $M$. Then, property (\emph{ii}) implies
$\sum_{k\in\mathcal{K}}S_k=\sum_{\nu \in M} E_\mu E_\nu=E_\mu=S_j$
and thus $\sum_{k\in\mathcal{K}}\Gamma_{ik}=\Gamma_{ij}$ . As
another example, suppose that $S_i=E_\alpha$ and $S_j=E_\beta
E_{\alpha'}$ with $\alpha,\alpha'\in X$ and $\alpha\neq\alpha'$.
Then, using successively properties (\emph{iii}),
(\emph{iv}) and (\emph{i}), we find $S^\dag_i S_j=E_\alpha E_\beta E_{\alpha'}=E_\alpha
E_{\alpha'}E_\beta=0$, and thus $\Gamma_{ij}=0$. The
conditions \eqref{lin2} are obtained by making use of
\eqref{qprob} in \eqref{gamma} and relate the entries of the
matrix $\Gamma$ to the specific set of probabilities
$P_{\alpha\beta}$ under consideration. Finally, to establish
\eqref{sdpcond}, remember that an $n\times n$ matrix $\Gamma$ is
semidefinite positive if and only if $v^\dag \Gamma v\geq 0$ for
all $v\in \mathbb{C}^n$. Expanding this expression,
we get \beq\begin{split}
v^\dag \Gamma v&=\sum_{i,j} v_i^* \text{tr}\left(S_i^\dag S_j \rho\right)v_j\\
&=\text{tr}\Big[\big(\sum_i v_i S_i\big)^\dagger \big(\sum_j v_j S_j\big)\,\rho\Big]\geq 0\,,
\end{split}\eeq
since $\rho$ is a positive operator.

We thus just showed that for any quantum distribution $P_{\alpha\beta}$, there necessarily exists for each set $\mathcal{S}$ a matrix $\Gamma$ satisfying the linear constraints \eqref{lin1} and \eqref{lin2}
and the condition of positivity \eqref{sdpcond}. If for some
$\mathcal{S}$ it is not possible to find a matrix $\Gamma$
satisfying these properties, we can therefore conclude that the
correlations characterized by the distribution $P_{\alpha\beta}$
cannot be reproduced through local measurements on a quantum
state. Determining if there exists a matrix $\Gamma$ satisfying conditions \eqref{lin1}, \eqref{lin2}, and \eqref{sdpcond} amounts to find a positive semidefinite matrix satisfying a set of linear constraints and is a typical instance of
semidefinite programming (see \cite{sdp} for a review of SDP). All the techniques developed in this context can thus be applied to evaluate our conditions.

To give a concrete example of our method, consider the case where
$\mathcal{S}=\{E_\alpha\}\cup\{E_\beta\}$ is simply the set of projectors of Alice and Bob. Suppose that there are $m$
different measurement outcomes $\alpha=1,\ldots,m$ for Alice and
$m$ different outcomes $\beta=m+1,\ldots,2m$ for Bob. Then
$\mathcal{S}=\{E_1,\ldots,E_m,E_{m+1},\ldots,E_{2m}\}$ and
applying the construction defined by \eqref{gamma}, we find that $\Gamma$ is a
$2m\times 2m$ matrix of the form \beq\label{exgamma}
\Gamma=\begin{pmatrix}
Q &P\\
P^T&R
\end{pmatrix}\,,
\eeq where the submatrix $P$ is simply the $m\times m$ table of
probabilities with entries $P_{\alpha\beta}$, and the submatrices
$Q$ and $R$ satisfy
\begin{equation*}\begin{aligned}\label{q}
Q_{\alpha\alpha}&=P_{\alpha},\qquad Q_{\alpha\alpha'}=0\quad (\alpha,\alpha'\in X \text{ and }\alpha\neq\alpha')\,,\\
R_{\beta\beta}&=P_\beta,\qquad R_{\beta\beta'}=0\quad(\beta,\beta'\in Y \text{ and } \beta\neq\beta')\,,
\end{aligned}\end{equation*}
where $P_\alpha=\sum_{\beta\in Y}P_{\alpha\beta}$ and
$P_{\beta}=\sum_{\alpha\in X}P_{\alpha\beta}$ are the marginal
probabilities for Alice and Bob, respectively. The form of the
matrix \eqref{exgamma} is defined by the linear constraints
\eqref{lin1} and \eqref{lin2}. The only entries of $\Gamma$ which
are not determined by these constraints are the entries
$Q_{\alpha\alpha'}$ with $\alpha\in X$ and $\alpha'\in X'$
belonging to different measurements of Alice ($X\neq X'$), and the
entries $R_{\beta\beta'}$ with $\beta\in Y$ and $\beta'\in Y'$
belonging to different measurements of Bob ($Y\neq Y'$ ). In a quantum scenario, these entries would correspond to non-commuting measurements performed on each subsystem and would thus be unobservable.
Nonetheless, if the correlations $P_{\alpha\beta}$ have a quantum origin it is possible to assign values to these undetermined entries such that the overall matrix \eqref{exgamma} is positive semidefinite, in accordance with \eqref{sdpcond}. As mentioned earlier, semidefinite programming can be used
to determine if the matrix \eqref{exgamma} can be completed in
such a way.

\paragraph{A hierarchy of conditions.}
We have shown how to design tests that distinguish quantum from non-quantum correlations. Each set $\mathcal{S}$ of operators used in our construction yields a different condition, and the choice of a particular $\mathcal{S}$ may thus seem arbitrary. But in fact not all conditions built according to our instructions are independent. Moreover, they can be organized in a hierarchical structure, such that they can all be checked in a systematic way.

To see this point, first note the easily established fact that if every operator in a set $\mathcal{S}$ can be
written as a linear combination of operators in another set
$\mathcal{S}'$, then the conditions obtained from $\mathcal{S}'$
are at least as constraining as the one obtained from
$\mathcal{S}$. By this we mean that if there exists a matrix $\Gamma'$ satisfying the constraints \eqref{lin1}, \eqref{lin2}, \eqref{sdpcond} associated to $\mathcal{S}'$, then there also exists a matrix $\Gamma$ satisfying the corresponding constraints associated to $\mathcal{S}$.
The second observation is that the set $\mathcal{T}_m=\{E_{\mu_1}\ldots E_{\mu_m}\}$ of all possible products of $m$ projectors generates by linear combinations all the operators that are linear combinations of products of $m'$ projectors, with $m'\leq m$.

Using these two properties, it is possible to remove the apparent arbitrariness on the choice of the set $\mathcal{S}$, and check all the conditions that can be built using our method in a comprehensive way. Start first with the condition based on the set $\mathcal{T}_1=\{E_\alpha\}\cup\{E_\beta\}$, which consists only of the projectors of Alice and Bob, and which we denote shortly as $\mathcal{T}_1=\{E_\mu\}$. If this condition is satisfied, consider the bigger set $\mathcal{T}_2=\{E_\mu
E_\nu\}$ consisting of all products of two projectors. If the corresponding test is passed, move to $\mathcal{T}_3=\{E_\mu E_\nu E_\tau\}$, and so on. Since the elements of $\mathcal{T}_{n-1}$ can be generated by linear combinations of the elements of $\mathcal{T}_n$, each successive test is at least as good as the previous ones. The sets $\mathcal{T}_n$, then, define a hierarchy of necessary conditions for testing the quantum origin of $P_{\alpha\beta}$. Note that if a test fails at some point in this hierarchy, we can immediately conclude that the correlations under consideration are not-quantum and there is no need to proceed with the successive tests.
The condition based on the matrix \eqref{exgamma} corresponds to the first test in this infinite hierarchy. As an illustration, we now present two applications of our method.

\paragraph{Application 1.} The first example involves two measurements for Alice, $X=1,2$, and two for Bob, $Y=3,4$, where each measurement yields one out of two outcomes, $+1$ or $-1$. This situation is thus
characterized by sixteen probabilities $P_{({\pm X})({\pm Y})}$,
to which we can associate eight projectors $E_{\pm M}$
($M=1,\ldots,4$). Alternatively, one can characterize this scenario by specifying the correlation functions $C_{XY}=\sum_{a,b} ab\, P_{(aX)(bY)}$ and the marginal quantities $C_X=\sum_{a}a\,P_{aX}$ and $C_Y=\sum_{b}b\,P_{bY}$ for each measurement. The first test in the hierarchy corresponds to the set $\mathcal{T}_1=\{E_{\pm
1},\ldots,E_{\pm 4}\}$, or, equivalently, to the set $\mathcal{S}=\{I,\sigma_1,\ldots,\sigma_4\}$, where $\sigma_{\!\scriptscriptstyle{M}}=E_{+M}-E_{{-}M}$, which is linearly equivalent to $\mathcal{T}_1$. The corresponding $5\times 5$ matrix $\Gamma$ is \beq
\Gamma=\begin{pmatrix}
1 & C_1 & C_2 & C_3 & C_4\\
& 1& u& C_{13} & C_{14}\\
& & 1 & C_{23} & C_{24}\\
& & & 1 & v \\
& & & & 1
\end{pmatrix}\,,
\label{gamma2}
\eeq
where we have only given the upper triangular part of $\Gamma$ since it
is hermitian. The parameters $u,v$ correspond to entries that are
not determined by our construction; but if the correlations represented by the quantities $\{C_X,C_Y,C_{XY}\}$ are quantum, it is possible to find
values for $u$ and $v$ such that the matrix \eqref{gamma2} is
semidefinite positive.

Note that a necessary condition for the matrix \eqref{gamma2} to be semidefinite positive is that the bottom-right  $4\times 4$ submatrix  satisfies
\beq
\bar\Gamma=\begin{pmatrix}
1& u& C_{13} & C_{14}\\
& 1 & C_{23} & C_{24}\\
& & 1 & v \\
& & & 1
\end{pmatrix}\succeq 0\,.
\label{gammacorr}
\eeq
This condition for the correlation functions $C_{XY}$ was already introduced by Landau \cite{landau} and Wehner \cite{wehner}. It is shown in \cite{landau} that there are values $u$ and $v$ such that \eqref{gammacorr} holds if and only if the correlations $C_{XY}$ satisfy the inequality
\beq\label{asin} |\asin C_{13} + \asin
C_{14} + \asin C_{23} - \asin C_{24}|\leq \pi\,, \eeq and the
three other ones obtained by permutation of the measurements. These inequalities have also been derived from a
different perspective by Tsirelson \cite{tsir2} and Masanes
\cite{masanes}.
%Actually, it follows from the characterization of quantum correlations
%given by Tsirelson \cite{tsir} that this condition is
%not only necessary but also sufficient for the correlation functions.

The correlation functions $C_{XY}$, however, only provide partial information about the full probability distribution. %In particular there exists non-quantum probability distributions whose correlation functions satisfy (\ref{asin}).
Our construction \eqref{gamma2}, on the other hand, deals with the full probability distribution, including the marginals $C_{X}$ and $C_{Y}$. Using the same techniques as used in \cite{landau} to derive (\ref{asin}) from \eqref{gammacorr}, it can be shown that the condition that the matrix \eqref{gamma2} is semidefinite positive is equivalent to the inequality
\beq\label{asin2} |\asin D_{13} + \asin D_{14} + \asin D_{23} -
\asin D_{24}|\leq \pi \eeq where
%\beq
%D_{ij}=\frac{C_{ij}-C_iC_j}{\sqrt{(1-C_i^2)(1-C_j^2)}}\,,
%\eeq
$D_{ij}=(C_{ij}-C_iC_j)/\sqrt{(1-C_i^2)(1-C_j^2)}$,
and to the inequalities obtained from \eqref{asin2} by permutation of
the measurement choices. If we neglect the marginals by imposing
$C_M=0$ we recover \eqref{asin}. As a test
on the full distribution, however, our inequality is stronger than \eqref{asin} since there are probability distributions that satisfy \eqref{asin} but
which violate \eqref{asin2}. Yet, \eqref{asin2} is not a
sufficient condition for a full probability distribution to admit a quantum representation, as we
have examples of correlations that satisfy \eqref{asin2} but which fail the successive step in the hierarchy.

\paragraph{Application 2.}
By maximizing the violation of a Bell inequality over the set of
probability distributions satisfying one of our conditions, we
obtain an upper-bound on the violation of this inequality by
quantum mechanics (since such conditions are satisfied by every
quantum distributions). A Bell inequality is a linear combination
of the probabilities $P_{\alpha\beta}$, and since these
probabilities are related in a linear way to the entries of the
matrices $\Gamma$, obtaining such an upper-bound can be cast as a
SDP. Let us stress that this bound is an upper-bound on the \emph{global} maximum of the inequality since we do not suppose anything about the measurements and states except that they are quantum.

As an example, consider the CHSH inequality, which reads $C_{13}+C_{14}+C_{23}-C_{24}\leq 2$.
Maximizing this expression over all distribution satisfying the
criterion \eqref{gammacorr} presented in Application 1 corresponds to the SDP,
\beq\begin{aligned}\label{chsh}
\text{maximize}&\quad C_{13}+C_{14}+C_{23}-C_{24}\\
\text{subject to}&\quad
\bar\Gamma\succeq 0\,.
\end{aligned}\eeq
The solution to this optimization problem is $2\sqrt{2}$, as noted
by Wehner \cite{wehner}, and we thus recover the well-known
Tsirelson bound. More generally, for any given Bell inequality,
SDPs can be associated to each of the conditions $\mathcal{T}_1,
\mathcal{T}_2,\ldots$ of our hierarchy, the solutions of which
would yield a sequence $I_1\geq I_2 \geq ...$ of upper-bounds on
the quantum violation of the inequality. Note that a tight bound may already be reached after a finite number of such iterations, as
the CHSH example shows.

We have applied the approach just outlined to the Collins et al.
inequalities \cite{cglmp}. This family of inequalities involves
two measurement choices per party and $d$ outputs per measurement,
and can be viewed as a generalization of the CHSH inequality for
systems of dimension greater than two. In \cite{acin},
lower-bounds for the maximal violation of the Collins et al. inequalities
are given for $d=3,\ldots,8$ by exhibiting a particular set of
measurements and states of dimension $d \times d$
yielding high violations of the inequalities. These quantum states
have the particularity to be non-maximally entangled. For $d=3$,
the reported violation is $I_*=2.9149$, the local bound of the
inequality being $I_{\text{loc}}\leq 2$. We have numerically
solved the SDP corresponding to the first tests in our hierarchy using the SeDuMi Matlab toolbox \cite{sedumi}.
For $d=3$, the condition $\mathcal{T}_1$ yields the bound $I\leq
3.1547$, which is about $10\%$ higher than the violation reported
in \cite{acin}. The second condition $\mathcal{T}_2$, however,
yields the bound $I\leq I_*=2.9149$, proving that the
partially entangled state and the measurements described in
\cite{acin} are optimal. We have also solved the SDP for
$d=4,\ldots,8$. As for the $d=3$ case, the first tests in the
hierarchy are about $10\%$ above the values of
\cite{acin}, but the second tests give the same results as the
ones reported in \cite{acin}, demonstrating that these are the
optimal quantum violations.

\paragraph{Conclusion.}
The approach given in this work opens a new way to study the
correlations between two separate quantum systems. There are
several possible extensions of our technique, for instance to
systems of more than two parties, and many potential applications
of it, among others to study non-local properties of quantum
correlations, such as their monogamous character \cite{bkp}. Two questions that
remain open is whether the hierarchy of conditions that we have introduced is complete, in the sense that a set of correlations satisfies every condition in the hierarchy if and only if it
admits a quantum representation, and, if this is the case, whether it is in general necessary to check all the conditions in this infinite hierarchy or whether it is sufficient to stop after a finite number of steps.

\acknowledgments We acknowledge support by the Spanish project FIS2004-05639-C02-02 and the EU Qubit Applications Project (QAP), Contract Number 015848. AA acknowledges support from a Spanish ``Ramon y Cajal"
grant. MN is supported by the Fundaci\'on Ram\'on Areces.

%\bibliography{qcondition}

\end{document}